\documentclass{bmcart}

\usepackage{amsthm,amsmath}
\usepackage{graphicx}
\usepackage[utf8]{inputenc} 

\startlocaldefs
\endlocaldefs

\begin{document}

\begin{frontmatter}

\begin{fmbox}
\dochead{Research}


\title{Machine Learning for \\Detecting Malware in PE Files} 


\author[
  email={cdc104@miami.edu}   
]{\inits{C.C.}\fnm{Collin} \snm{Connors}}
\author[
  email={sarkar@cs.miami.edu}
]{\inits{D.S.}\fnm{Dillip} \snm{Sarkar}}


\end{fmbox}


\begin{abstractbox}

\begin{abstract} 
The increasing number of sophisticated malware poses a major cybersecurity threat.
Portable executable (PE) files are a common vector for such malware.
In this work we review and evaluate machine learning-based PE malware detection techniques. Using a large benchmark dataset, we evaluate features of PE files using the most common machine learning techniques to detect malware.
\end{abstract}



\end{abstractbox}
%

\end{frontmatter}



\section{Introduction}
\label{Introduction}
\emph{Mal}icious soft\emph{ware} (Malware) infects digital systems (computers, smartphones, networks, cloud servers, etc.), harming the systems or their users.   The degree of harm to a system varies from reduced performance to unavailability for use or stealing sensitive information; the degree of harm to the users varies from damaging their documents to stealing users' information while using an infected system.  Based on behavior and damage to an infected system, malware is classified as adware, backdoor, bot, downloader,  launcher, ransomware, rootkit, spyware,  Trojan, virus, and worm, etc. (see~\cite{SurveyOfMalwareDetectionApproaches2020Access,SurveyRiseOfMachineLearningForDetectionAndClassificationOfMalwareDetectionResearchDevelopmentEtc2020JNCA}, and references therein).
\par
Early forms of malware, commonly known as traditional malware, had characteristic signature strings, and they are identified by detecting those signatures.   However, soon a race started between malware developers and anti-malware software developers. Newer malware utilized sophisticated and advanced techniques; thus, they have no simple signature for detection;  their replicates have different appearances yet similar harmful capabilities.  
\par
Several methods were developed for analysis and mining of advanced malware to discover their underlying characteristic features~\cite{SurveyOfMalwareDetectionTechniquesBasedOnMachineLearning2019Journal}. 
\par
 In addition to malware signatures,  these mining tools analyzed  {DLL} function calls, hexadecimal sequences,  assembly instructions,  {PE} file header, or some combination of these features~\cite{SurveyRiseOfMachineLearningForDetectionAndClassificationOfMalwareDetectionResearchDevelopmentEtc2020JNCA, SurveyOfMalwareDetectionApproaches2020Access,AStateOfTheArtSurveyOfMalwareDetectionApproachesUsingDataMiningTechniques2018Journal,SurveyOfMalwareDetectionTechniquesBasedOnMachineLearning2019Journal}. The features obtained after the initial pass may be further processed before feeding them in a tool. For example,~ entropy and hashing trick are used in \cite{EMBER}. Note that the so called Hashing Trick is not a trick but rather a well studied dimension reduction technique with a solid theoretical foundation~\cite{FullyUnderstandingHashingTrick2018Conf}.

\par
Advanced malware is adaptive and persistent; they evade detection after they infect systems. Thus, before any software file enters a digital system, it ought to be evaluated for malignancy.  Software evaluation methods are categorized as static, dynamic, and hybrid~\cite{AStateOfTheArtSurveyOfMalwareDetectionApproachesUsingDataMiningTechniques2018Journal}; hybrid analysis utilizes both static and dynamic analysis.

\par 
Dynamic analysis of software is performed by executing it in a `\emph{sandbox}' (an isolated environment that mimics end-user operating environments), observing and recording its behavior.  The recorded behavior of the software is evaluated manually or by machine learning techniques for final classification~\cite{AnalysisOfMachineLearningTechniquesUsedinBehavior-BasedMalwareDetection2009Conf}. Dynamic analysis is prolonged, expensive, and time-consuming and thus is impractical for most cases. 
\par
Static analysis of software is performed by extracting features present in the software by analysis or mining and then classifying it based on the features obtained.  The features needed for classification depend on the tool to be used. Five categories of tools for static analysis are signature-based, behavior-based, heuristic-based, model checking-based, and machine learning based~\cite{SurveyOfMalwareDetectionApproaches2020Access}.
\par
The malware developers have access to all detection tools, and they swiftly modify their code to evade one or more of these tools. Thus, malware detection tools must be ready to identify newer variants or the arrival of a new malware category. Unsurprising, the race between malware developers and malware detection tool developers is expected to continue forever. Fortunately, \emph{machine learning} (ML) based tools often successfully identify newer variants and \emph{learn to detect new malware category, by \emph{incrementally} retraining the model after adding samples of newer variants and categories in the training dataset}.

\subsection{Machine Learning  for Malware Detection}
\label{sec:MalwareDetectionUsingMachineLearningMethods}
\par
Signature-based malware detection tools' malware detection rate decreased because advanced malware lacked signature strings.
Search for alternative detection techniques converged to ML-based tools. The first ML-based tools were reported in 2001~\cite{MalwareDetectionUsingMachineLearningFirstPaper2001Conf}. 
Since then, numerous ML-based models have been reported and they have been compared and contrasted in many survey papers.  The most recent and comprehensive of these are~\cite{SurveyRiseOfMachineLearningForDetectionAndClassificationOfMalwareDetectionResearchDevelopmentEtc2020JNCA,SurveyOfMalwareDetectionApproaches2020Access,SurveyMachineLearningTechniquesMalwareAnalysis2019JournalUCCI}.  
\par
Until recently, the ML-based models were developed using small datasets, and their validation did not use unseen \emph{future} malware.  The systems developed and evaluated using the EMBER dataset~\cite{EMBER} overcame both of these problems because the dataset is large, data is divided into subsets based on malware's observation time; the training dataset is from the \emph{past}.
 Moreover, the evaluation dataset is from the \emph{future}, that is,  each PE file in the  evaluation dataset was detected after  all PE files in the training dataset were detected.  Since the EMBER data set has a time-based partition, work reported 
in~\cite{EvaluatingPerformanceMaintenanceAndDeteriorationOverTimeOfMLbasedMalwareDetection2020Conf} evaluated deterioration over time.  
\par
In this work, first, we briefly describe the structure and elements of Portable Executable (PE) files,  review a benchmark dataset, and the extracted feature sets provided with the dataset~(see Table~\ref{tab:vdata} for details).  Also, we briefly describe ML algorithms that have been proposed and evaluated in the past, and then report the performance of  ML-based systems using {EMBER} dataset. Finally, we experimentally evaluate the contributions of the nine feature sets towards the detection of malware.

\section{Portable Executable Files}
\label{sec:PortableExecutableFiles}
\par
The portable executable (PE) file format is a windows based file format for executable files based on the UNIX Common Object File Format. The PE format contains instructions for the windows dynamic linker on loading the code stored in the file and necessary libraries to execute the code into memory. A few common PE file extensions include .exe, .dll, .sys, .cpl, and .tsp. A simplified layout of the Windows PE file is shown in Figure~\ref{fig:PEFileFormat}.

\par
PE files consist of two parts, the PE header, and the PE sections. The PE sections contain the raw data associated with the PE file. While sections can take any name there are a few common section names found in most PE files. Common sections include:
\begin{itemize}
	\item .text - which includes the file's code,
	\item .data - which includes initialized data used by the file,
	\item .rdata - which includes read-only data to be used by the file
	\item .bss -which contains uninitialized data.
\end{itemize}

\par
The PE header contains instructions on where libraries are located and how to load the file's code into memory. Most PE headers will include section headers, a DOS Header, and Optional Headers. The Optional Header is required for all executable files and contains the Data Directories where necessary libraries are stored, the Import table, which contains the functions imported by the PE file, and the export table, which contains functions that other PE files can import. 

\par
Because PE files can be run on most versions of windows, they are the most popular vector for malware. The malware detection site VirusTotal reported in 2016 that a plurality of files scanned, 47.8\%, were PE files \cite{KUMAR2019252}. Since the PE file format is a popular vector for malware, researchers have assembled datasets containing both malicious and benign PE file samples. 
\section{The EMBER Dataset}
\label{sec:TheEMBERDataset}

\par
The Elastic Malware Benchmark for Empowering Researchers (EMBER) dataset \cite{EMBER} is a dataset consisting of preprocessed malicious and benign PE files. The dataset contains 900,000 training samples and 200,000 testing samples. The training set consists of 300,000 samples of  malicious, benign, and unlabeled files, while the testing set consists of  100,000 samples of  malicious and benign PE files. 
\par
The EMBER dataset v1 was initially released in 2017, followed in 2018 with EMBER dataset v2. The most significant change is that EMBER v2 added a new feature set, data directory, that was not present in EMBER v1. The paper released with the dataset used v1 while our work uses v2; thus, some of our results may vary slightly from what was initially reported.
\par
EMBER consists of preprocessed PE files and software for vectorizing the feature sets found in each file. The length of each feature can be found in Table~\ref{tab:vdata}, which sums to 2381. Below is a brief description of these nine feature sets. For the convenience of referring to these features, we abbreviate each with two letters as shown within parenthesis.

\begin{enumerate}
		\item Byte histogram (BH)  - The histogram of Byte distribution in the file
    \item Byte-entropy histogram (BE)  - The histogram of the byte entropy in the file
    \item String information (ST) - Information about the strings found in the file
    \item General file information (GE) - General information about the file such as the size of the file
    \item Header information (HE) - Information found in the section header of the file such as number of sections and size of sections
    \item Section information (SE) - Information from each section, this feature set is generated by using the Hashing Trick~\cite{FullyUnderstandingHashingTrick2018Conf}
		\item Imported functions (IM) - The functions that the file calls, this feature set is generated by using the Hashing Trick on the library calls and then on the function calls
    \item Exported functions (EX) - The functions that the file exports, this feature set is generated using the Hashing Trick
		\item Data Directory (DD) - The directories where critical files needed to run the portable executable can be found. 
\end{enumerate}
\par
The Byte Histogram, Byte-Entropy Histogram, and String Information are not unique to the PE file format and could be used to make a more general malware detection model. The EMBER dataset includes code to vectorize the data. We will use the vectorized data in our models. The process for vectorization is described in detail in the EMBER paper \cite{EMBER}. Several feature sets use the so called `Hashing Trick' for feature extractions, the fact is that it is not \emph{a trick} at all; it is a well-studied dimension reduction technique with a solid theoretical foundation~\cite{FullyUnderstandingHashingTrick2018Conf}. 
\par
Table~\ref{tab:vdata} shows the description of each of the feature sets after vectorization. Some of the feature sets, such as sections, have a wide range, while others, such as byte histogram, have a narrow range. To account for this, all of our models use feature scaling to keep feature sets in similar ranges and prevent one feature set from dominating the models. Likewise, the length of each feature set varies. We will look at methods for reducing the number of feature sets to see if a smaller dataset results in better performance.
\begin{table}[htb]
\resizebox{\textwidth}{!}{%
\begin{tabular}{llll}
 Feature &Abbr.&Feature&Abbr.\\ \hline
Histogram     & BH     &   Sections      & SE\\
ByteEntropy   & BE    & Imports       & IM\\
Strings       & ST     & Exports       & EX \\
General       & GE&DataDirectory & DD   \\
Header        &HE  &&\\   
\vspace{0.5pt}        
\end{tabular}%
}
\caption{Abbreviations used for each feature set}
\label{tab:FeaturesAbbr}
\end{table}
\begin{table}[htb]
\resizebox{\textwidth}{!}{%
\begin{tabular}{llrrr}
     Feature         & Data  &\multicolumn{1}{c}{Minimum } & \multicolumn{1}{c}{Maximum } & \multicolumn{1}{c}{Size} \\
		            & Type &\multicolumn{1}{c}{value} & \multicolumn{1}{c}{value} & \multicolumn{1}{l}{} \\
 \hline
		
BH    & Float     & 0                       & 0.9999558               & 256                      \\
BE   & Float     & 0                       & 0.9997657               & 256                      \\
ST       & Int+      & 0                       & 61745440                & 104                      \\
GE       & Int+      & 0                       & 8589935000              & 10                       \\
HE       & Int       & -2                      & 4294967300              & 62                       \\
SE      & Int       & -4294967300             & 6167139300              & 255                      \\
IM       & Int       & -882858                 & 83917                   & 1280                     \\
EX       & Int       & -2288                   & 308                     & 128                      \\
DD & Int+      & 0                       & 4294967300              & 30       \\
\vspace{0.5pt}               
\end{tabular}%
}
\caption{Data type, minimum and maximum values, and size of each feature set extracted from a PE file.}
\label{tab:vdata}
\end{table}
\section{ML-Based Malware Detection}
\label{sec:LearningBasedMalwareDetection}
After a brief introduction to  categories of ML algorithms, we review ML-algorithms that have been used for developing  ML-based malware detection systems.
\subsection{Categories of ML-algorithms}
\label{sec:CategoriesOfMLAlgorithms}
\par
A Machine Learning (ML) algorithm (Readers not familiar with ML algorithms can read them from one of the many ML books, for example~\cite{MachineLearningTextMitchell97}) creates an intelligent system from sample data, known as \emph{training data},  presented to it;  the created system is evaluated using  \emph{unseen} sample data, known as \emph{evaluation data}.
A sample datum marked with the identity of the class where it belongs is known as \emph{labeled datum}. For simplifying the discussion, we assume that a software file is either malicious or benign and thus, a software's label is either malware or benign.  Unless otherwise stated,  we assume that a trained system classifies an input file as either malware or benign.
\par
 Based on the types of training data used, ML algorithms are divided into three categories: \emph{unsupervised}, \emph{supervised}, and \emph{semi-supervised}.   
An unsupervised ML algorithm is trained with unlabeled data. During training, the ML algorithm finds parameter values of the model for clustering the training data into two clusters: one cluster is for malware, and another cluster is for benign software; simply stated, an unsupervised ML algorithm uses a distance measure to determine cluster membership (see \cite{AStateOfTheArtSurveyOfMalwareDetectionApproachesUsingDataMiningTechniques2018Journal} for different distance measures that have been used for training malware detection systems).  When an unseen datum is presented to a trained unsupervised system, it classifies it to the cluster whose distance is minimum.

\par
Algorithms in the supervised category are presented with \emph{labeled} training data; that is, each sample has a class label.  We use supervised models in this work. The semi-supervised category of algorithms is presented with some labeled and unlabeled data; this category's definition varies in the literature. Most of the successful ML-based malware detection tools are trained with \emph{extensive} training datasets.

\subsection{Algorithms Used for ML-based  Detection Tools}
\label{sec:HistoryOfMalwareDetectionUsingML}
\par
First ML-based malware detection  tools, proposed in 2001, used 
\emph{Rule-based classifier} and \emph{Naive Bayes} learning algorithms~\cite{MalwareDetectionUsingMachineLearningFirstPaper2001Conf}; subsequently, potentially all machine learning algorithms have been used  for creating ML-based  malware detection tools~\cite{AStateOfTheArtSurveyOfMalwareDetectionApproachesUsingDataMiningTechniques2018Journal,
SurveyMachineLearningTechniquesMalwareAnalysis2019JournalUCCI,
SurveyOfMalwareDetectionApproaches2020Access,
SurveyOfMalwareDetectionTechniquesBasedOnMachineLearning2019Journal,
SurveyRiseOfMachineLearningForDetectionAndClassificationOfMalwareDetectionResearchDevelopmentEtc2020JNCA}.  
Among the survey papers,~\cite{SurveyRiseOfMachineLearningForDetectionAndClassificationOfMalwareDetectionResearchDevelopmentEtc2020JNCA} is most comprehensive; it has tabulated ML algorithms in three tables --- one each for supervised, semi-supervised, and unsupervised categories. Also, the tables have sources of the datasets,  their sizes, and features of the datasets used for training and evaluation of the malware detection systems.   
\par
From these tables, we find that very few datasets were in the public domain. Many of them are no longer available today, and most of them are too small for developing any real-world malware detection system.  
Moreover, with one exception~\cite{DeepNeuralNetworkBasedMalwareDetectionUsingBinaryFeatures2015Conf} we suspect that these datasets  \emph{did not} include the discovery date of the malware; thus, the developed models were not evaluated for endurance with evolving malware variants and new varieties of malware.

\subsection{Previously used ML-algorithms}
\label{sec:ListOfMLAlgorithmsUsed}
\par
Machine learning algorithms used for training malware detection systems include Artificial Neural Networks, Bayes classifier, Bayesian Networks, Belief Propagation, different Clustering, Decision trees (DT), Gradient Boosting Decision Trees (GBT),  $k$-Nearest Neighbors ($k$-NN / KNN),  Linear Discriminant Analysis  (LDA), Logistic Model Tree, Logistic Regression, Multiple Kernel Learning, Naive Bayes (NB), Prototype-based Classification, Quadratic Discriminant Analysis (QDA), Random Forest (RF),  Rule-based classifiers,      Self-Organizing Maps,  and Support Vector Machine (SVM). 
 Interestingly, 75\% of the systems were developed using four ML algorithms ---  DT (31\%), SVM (29\%),  NB (10\%), and RF (5\%)~\cite{AStateOfTheArtSurveyOfMalwareDetectionApproachesUsingDataMiningTechniques2018Journal}. 
\par
Most recently, ANN, including Deep Neural Networks (DNN), and a variant of  GBT known as Light Gradient Boosting Machine (LGBM), have shown good performance when trained and evaluated using a large dataset~\cite{EMBER}, {EMBER} (described earlier).  

\par 
ML-based systems developed and evaluated earlier have provided significant insights, and these insights are valuable. However, the reported results cannot be compared with each other because they were developed and evaluated using myriad and relatively small datasets. Availability of the large EMBER
dataset makes a compelling case for developing and evaluating all those ML-based systems again. 
\par 
To that end, using Version 0.23.2 of Scikit-learn: Machine Learning in Python  package~\cite{scikit-learn} and the EMBER dataset, we developed and evaluated malware detection systems.
 
\section{Performance of ML-based Systems Developed  using {EMBER} dataset}
\label{sec:MLBasedSystemsDevelopedAndEvaluatedUsingEMBERDataset}

Our study includes LGBM (original and normalized data), RF, LG, LDA, KNN, QDA, SVM,
and ANN/DNN ML-algorithms;  all 9 feature sets of the {EMBER} dataset were used to develop these systems. 
\par
Their malware detection rates are shown in Table~\ref{tab:StatisticalModels}.  (The program codes will be made available from our GitHub once the paper is accepted for publication.)   Because the paper describing the EMBER dataset (benchmark)  reported that LGBM has the best detection rate, we consider the performance of LGBM  as the baseline performance. The table entries are sorted from the highest accuracy to the lowest accuracy (except ANN/DNN).  The accurate detection rate for the RF is 1.67\% below LGBM, which is not a surprise because LGBM  is an optimized version of RF. The detection accuracy of LG, LDA, and KNN systems is 87.37\%, 85.82\%, and 82.23\%, respectively. The systems developed using SVM, and QDA performed very poorly --- 54.18\% and 50.29\%, respectively; these are almost identical to a coin toss. 

\begin{table}[]
\resizebox{\textwidth}{!}{%
\begin{tabular}{lccccc}
              & \multicolumn{1}{c}{ACC} & \multicolumn{1}{c}{AUC} & \multicolumn{1}{c}{Presision} & \multicolumn{1}{c}{Recall} & \multicolumn{1}{c}{F1} \\ \hline
LGBM \\Baseline & \textbf{0.9363}         & \textbf{0.9845}         & 0.9244                        & 0.9504                     & \textbf{0.9372}        \\
LGBM \\Normalized     & 0.9351 & 0.9844 & 0.9267          & 0.9450          & 0.9357 \\
RandomForest        & 0.9196 & 0.9196 & 0.9208          & 0.9183          & 0.9195 \\
Logistic \\Classifier & 0.8737 & 0.8737 & 0.8394          & 0.9243          & 0.8798 \\
LDA                 & 0.8582 & 0.8582 & 0.8175          & 0.9222          & 0.8667 \\
KNN                 & 0.8223 & 0.8223 & 0.7979          & 0.8632          & 0.8293 \\
SVM                 & 0.5418 & 0.5418 & 0.5222          & \textbf{0.9853} & 0.6826 \\
QDA                 & 0.5029 & 0.5029 & \textbf{0.9770} & 0.0059          & 0.0118 \\
ANN/DNN             & 0.9522 & 0.9815 & 0.9592          & 0.9445          & 0.9518\\
\vspace{0.5pt}          
\end{tabular}%
}
\caption{Performances of ML-based systems on the EMBER dataset, from the highest to the lowest ACCs. The AUC is reported for a false-positive rate less than 1\%.}
\label{tab:StatisticalModels}
\end{table}

\par
As can be seen from the last row of the Table~\ref{tab:StatisticalModels}, the performance of our proposed ANN/DNN system is better than the performance of the baseline LGBM-based system. We describe our ANN/DNN architecture and compare its performances with the performances of the baseline system.

\subsection{Neural Network Approach to Malware Detection}
\label{sec:NeuralNetworkApproachToMalwareDetection}
\par
We extensively evaluated various ANN/DNN architectures. We found that large deep networks do not increase detection rate above the baseline LGBM, but they needed significantly higher training time and slightly higher detection time. We found a simple ANN architecture  (shown in Fig.~\ref{fig:Network_Architecture}) that outperformed all systems. As can be seen from the figure, it consists of three dense layers of 512, 128, and 8 neurons with \emph{tanh} activation functions. The output layer consists of a two-unit $softmax$ layer. A batch normalization layer is used before feeding the inputs to the network; also, we use another batch normalization layer between the 2nd and 3rd dense layers. Performance of this ANN-based system is shown in the last row; we can see that it improves accurate detection rate  from 93.363\% to 95.22\% -- a 1.27\% increase;  but if one considers the reduction of malware detection rate, the ANN-based system reduces
inaccurate detection rate to 4.78\% from LGBM's 6.05\% --- a reduction of 20.99\%.

\section{Evaluation of the Nine Feature Sets of the EMBER Dataset}
\label{sec:NeuralNetworkApproachToMalwareDetection2}
\par
Multiple features from PE files have been used in~\cite{DeepLearningApproachToFastFormatAgnosticDetectionOfMaliciousWebContent2018Sophos} and~\cite{EMBER}. We were interested to see how many contributions each of these features make towards the detection rates.  Because ANN and LGBM based systems have two top detection rates, we only report our observations for those systems. It is important to note that the length of the input to a system is the sum of lengths of the feature sets used to train it. 

\subsection{Single Feature Set}
\label{sec:SingleFeature}
Performance of the ANNs and LGBMs with only one feature set is shown in Table~\ref{tab:One}. First, two columns of the table are for accuracies (ACCs) and  AUCs (AUC stands for area under the curve and is the area under the receiver operator characteristic (ROC) curve.) at a false-positive rate less than 1\%, and the 3rd and the 4th columns are for those of the LGBM. While most of the features show similar detection rates, there are a couple of notable differences.

\begin{table}[htb]
\resizebox{0.9\columnwidth}{!}{%
\begin{tabular}{l|rr|rr}
 & \multicolumn{2}{l}{ANN Performance}               & \multicolumn{2}{l}{LGBM Performance}              \\
 & \multicolumn{1}{l}{ACC} & \multicolumn{1}{l}{AUC} & \multicolumn{1}{l}{ACC} & \multicolumn{1}{l}{AUC} \\ \hline
BH & \textbf{0.8949} & \textbf{0.9567} & 0.8657          & 0.9465          \\
BE   & 0.8912          & 0.9566          & 0.8529          & 0.9368          \\
ST        & 0.8859          & 0.9511          & 0.8645          & 0.9462          \\
IM        & 0.8662          & 0.9338          & 0.8300          & 0.9172          \\
SE      & 0.8648          & 0.9359          & \textbf{0.8757} & \textbf{0.9522} \\
GE        & 0.7316          & 0.7924          & 0.8285          & 0.9243          \\
HE         & 0.7312          & 0.8088          & 0.8102          & 0.9110          \\
EX        & 0.5571          & 0.5852          & 0.5531          & 0.5610          \\
DD & 0.5050          & 0.4852          & 0.8473          & 0.9356      \\ 
\vspace{0.5pt}  
\end{tabular}%
}
\caption{The results of our ANN and the LGBM systems with only a single feature set at a time. The AUC is for a false-positive rate less than 1\%.}
\label{tab:One}
\end{table}

\par
The ANN-based system has the highest detection rate of 89.49\% with  BH (byte histogram), and that for the LGBM based system is 87.57 with SE (sections). For most \emph{single} feature sets, we observe that performances of ANN and LGBM systems are close to each other, with the only exception: DD (data directory); ANN-based system performed very poorly (at 50.50\%), but LGBM based system achieved an 84.73\% accuracy.
Another notable difference is HE (header information), where LGBM had a detection rate of 81.02\% vs. 73.12\% for the ANN.

\par Next we present our observations for combinations of two feature sets at a time.

\begin{table}[htbp]
\resizebox{.95\columnwidth}{!}{%
\begin{tabular}{l|cc|cc}
                         & \multicolumn{2}{l}{ANN Preformance}               & \multicolumn{2}{l}{LGBM Preformance}  \\
                         & \multicolumn{1}{l}{Acc} & \multicolumn{1}{l}{AUC} & \multicolumn{1}{l}{Acc} & \multicolumn{1}{l}{AUC} \\ \hline
BH\_SE & \textbf{0.9342}      	 & \textbf{0.9745}      	 & 0.9129              		 &  0.9737 \\
BH\_IM  & 0.9313                  & 0.9751                  & 0.9064                  &  0.9670 \\
BE\_IM  & 0.9292                  & 0.9702                  & 0.9044                  &  0.9686 \\
BE\_SE & 0.9275                  & 0.9724                  & 0.8962                  &  0.9683 \\
ST\_SE        & 0.9255               & 0.9718               & \textbf{0.9161}      &  \textbf{0.9762} \\
$\vdots$&$\vdots$&$\vdots$&$\vdots$&$\vdots$\\
GE\_EX         & 0.7372                  & 0.8031                  & 0.8315                  &  0.9233 \\
HE\_EX          & 0.7304                  & 0.8087                  & 0.8107                  &  0.9098 \\
HE\_DD   & 0.7274                  & 0.8083                  & 0.8621                  &  0.9489 \\
GE\_DD  & 0.7203                  & 0.7841                  & 0.8796                  &  0.9539 \\
EX\_DD  & 0.6157                  & 0.6367                  & 0.8470                  &  0.9341\\ \hline
\vspace{0.5pt}
\end{tabular}%
}
\caption{The results of our ANN and the LGBM systems with combinations of two feature sets. The AUC is reported for a false-positive rate less than 1\%.}
\label{tab:Pairs}
\end{table}

\subsection{Combination of Two Feature Sets}
\label{sec:CombinationOfTwoFeatures}
Out of nine feature sets, one can choose two features simultaneously in 36 distinct possible ways.  We evaluated all 36 possible combinations of two feature sets, but we show the top five and bottom five pairwise combinations to conserve space and reduce clutter. 
It is important to note that the set of included two-feature set combinations has the best performances of both ANN  and LGBM  based systems.  Our ANN-based system has the highest malware detection rate of 93.42\% with the feature sets BH (byte histogram) and SE( Section information); the LGBM based system has the highest malware detection rate of  91.61\%  with the features ST (String information) and SE --- both share the feature set SE.
\par
While differences in malware detection rates in the top five observations vary a little, the picture for the bottom five is different;  our ANN-based systems' detection rate is drastically lower than that of the LGBM based systems' detection rate.  Our ANN-based systems have the lowest detection rate of 61.57\% with feature sets EX (exports) and  DD (Data directories), but the lowest detection rate for the LGBM based system is 84.70\% --- a difference of more than 23\%; fortunately, there is no good reason to choose the worst-performing combinations.
\par
The top five pairwise combinations include features BH, BE, SE, IM, and ST. Looking back at performances of single feature set based systems, these five features produced the top five systems (see Table~\ref{tab:One}). Comparing Tables~\ref{tab:One}~and~\ref{tab:Pairs}, we can find that BH and BE produced two top-performing systems with single feature sets, but the combination of these two feature sets did not produce any of the top-performing five systems for ANNs. 
\par
To create and evaluate systems with three or more features, we will consider only BH, BE, SE, IM, and ST. Next, we discuss the performances of systems created with three feature sets.

\begin{table}[htbp]
\resizebox{\textwidth}{!}{%
\begin{tabular}{l|rr|rr}
                                       & \multicolumn{2}{l}{ANN Preformance} & \multicolumn{2}{l}{LGBM Preformance} \\
                                  & \multicolumn{1}{l}{Acc} & \multicolumn{1}{l}{AUC} & \multicolumn{1}{l}{Acc} & \multicolumn{1}{l}{AUC} \\ \hline
BH\_SE\_IM & \textbf{0.9471}         & \textbf{0.9810}         & 0.9262                  & 0.9800                  \\
BE\_SE\_IM        & 0.9458           & 0.9799           & 0.9228            & 0.9770           \\
ST\_SE\_IM             & 0.9447           & 0.9803           & \textbf{0.9276}   & \textbf{0.9809}  \\
BH\_BE\_SE & 0.9370           & 0.9754           & 0.9181            & 0.9766           \\
BH\_ST\_SE      & 0.9352           & 0.9749           & 0.9201            & 0.9784           \\
BE\_ST\_SE        & 0.9345           & 0.9740           & 0.9194            & 0.9781           \\
BE\_ST\_IM         & 0.9340           & 0.9729           & 0.9154            & 0.9740           \\
BH\_BE\_IM  & 0.9334           & 0.9749           & 0.9134            & 0.9728           \\
BH\_ST\_IM       & 0.9317           & 0.9735           & 0.9118            & 0.9738           \\
BH\_BE\_ST  & 0.9164           & 0.9630           & 0.8983            & 0.9639          \\
\vspace{0.5pt}
\end{tabular}%
}
\caption{The results of our ANN and the LGBM systems with combinations of three feature sets. Only Byte Histogram, Byte Entropy, Strings, Sections, and Imports were tested. The AUC is reported for a false-positive rate less than 1\%.}
\label{tab:Triplets}
\end{table}

\subsection{Combination of Three Features}
\label{sec:CombinationOfThreeFeatures}
Performances of systems created with three out of five features are shown in table Table~\ref{tab:Triplets};  we have included results for all ten possible combinations.   Among all the LGBM based systems created and evaluated with combinations of three features, the system created with ST, SE, and IM has the highest malware detection rate of 92.75\% (this is less than 1\% lower than the baseline system with 93.63\% detection rate).  The most interesting observation is the performances of the ANN-based systems. The top four systems outperform the baseline system's performance, and the best detection rate of 94.71\% is obtained by our ANN-based system created with BH, SE, and IM. Another interesting observation is that SE is the common feature among all the top six ANN-based systems.

\begin{table}[htbp]
\resizebox{\textwidth}{!}{%
\begin{tabular}{l|rr|rr}
                                                & \multicolumn{2}{l}{ANN Preformance} & \multicolumn{2}{l}{LGBM Preformance} \\
                                                         & \multicolumn{1}{l}{Acc} & \multicolumn{1}{l}{AUC} & \multicolumn{1}{l}{Acc} & \multicolumn{1}{l}{AUC} \\ \hline
ALL                                             & \textbf{0.9522}  & \textbf{0.9815}  & \textbf{0.9363}   & \textbf{0.9845}  \\
BE\_ST\_SE\_IM        & 0.9500           & 0.9808           & 0.9310            & 0.9827           \\
BH\_BE\_ST\_SE\_IM & 0.9497                  & 0.9818                  & 0.9339                  & 0.9833                  \\
BH\_ST\_SE\_IM      & 0.9470           & 0.9797           & 0.9332            & 0.9830           \\
BH\_BE\_SE\_IM & 0.9464           & 0.9817           & 0.9297            & 0.9808           \\
BH\_BE\_ST\_SE & 0.9381           & 0.9745           & 0.9234            & 0.9799           \\
BH\_BE\_ST\_IM  & 0.9362           & 0.9749           & 0.9183            & 0.9761       \\
\vspace{0.5pt}   
\end{tabular}%
}
\caption{The results of our ANN architecture and the LGBM against quadruplets of feature sets, quintuplets of feature sets, and all of the feature sets. Only Byte Histogram, Byte Entropy, Strings, Sections, and Imports were tested for the quadruplets and quintuplets. }
\label{tab:FourUp}
\end{table}

\subsection{Combinations of Four or More Feature Sets}
\label{sec:CombinationsOfFourOrMoreFeatures}
Results of systems created with four or more features are shown in Table~\ref{tab:FourUp}. 
As discussed and analyzed earlier, our ANN-based systems with all features have a malware detection accuracy of 95.22\% and significantly outperform the baseline systems created with LGBM. Another observation is that our ANN-based systems have outperformed the baseline LGBM based system.
\section{Conclusion and Future Work}
\label{sec:ConclusionAndFutureWork}
Availability of the large {EMBER} dataset~\cite{EMBER} has opened the door for creating and evaluating malware detection systems. The dataset includes more than just the PE files; it includes the program code necessary to extract nine features from each PE file and the program code for LGBM; we have used these programs to obtain the baseline results we have reported. The program code for extracting features from each PE file produces a long vector, concatenating all nine features. We separated each of those features for our study reported here. 
\par 
We have seen that ANN-based systems can outperform other ML-based systems; we have shown that all features do not contribute equally. Even when two features perform well individually, they may not have any synergy; together, they may not produce a system that performs much better than the systems created with each feature individually; our study revealed that Byte histogram and  Byte entropy are such a pair of features.  
\par 
Another interesting observation is that two ML algorithms may create systems with the same set of features, but their performances vary widely; we have seen that for our ANN-based and LGBM based systems.



\begin{backmatter}

\section*{Declarations}

\subsection*{Availability of data and materials}
Our models are available upon request.

\subsection*{Competing interest}
The authors declare that they have no competing interests.

\subsection*{Funding}
None.

\subsection*{Authors' contributions}
The first author conceived the idea of the study and wrote the code and the paper; both authors discussed the results and revised the final manuscript. Both authors read and approved the final manuscript.

\subsection*{Acknowledgements}
Not applicable.


\bibliographystyle{plain} 
\bibliography{ref}      




\section*{Figures}
\begin{figure}[h!]
	\centering
		\includegraphics[height=.5\textheight]{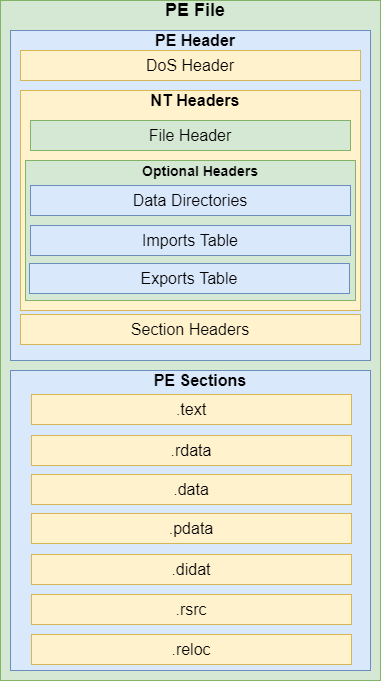}
	\caption{A simplified view of the notepad.exe PE file to highlight the anatomy of the PE file format.}
	\label{fig:PEFileFormat}
\end{figure}

\begin{figure}[h!]
	\centering
		\includegraphics[height=.5\textheight]{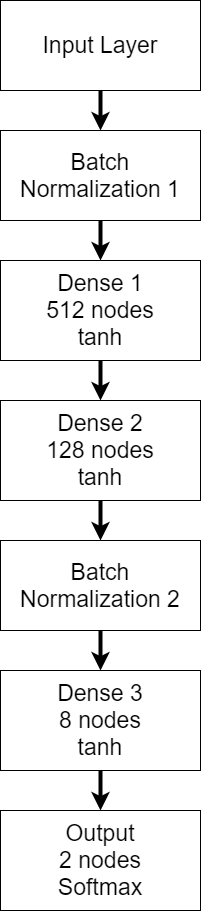}
	\caption{A simple but effective ANN architecture for malware detection,}
	\label{fig:Network_Architecture}
\end{figure}

\end{backmatter}
\end{document}